\documentclass[apjl]{emulateapj}
\usepackage{psfig,amsfonts,amsmath,graphicx,natbib,apjfonts}
\def\ra#1#2#3{#1$^{\rm h}$#2$^{\rm m}$#3$^{\rm s}$}
\def\dec#1#2#3{#1$^\circ$#2$'$#3$''$}

\def\swift{{\it Swift}}
\def\chandra{{\it Chandra}}
\def\grb{GRB\,120804A}

\def\cfa{1}
\def\war{2}
\def\inaf{3}
\def\psu{4}
\def\lei{5}
\def\mpi{6}
\def\dcc{7}
\def\gem{8}

\begin{document}

\title{The Afterglow and ULIRG Host Galaxy of the Dark Short
GRB\,120804A}

\author{
E.~Berger\altaffilmark{\cfa},
B.~A.~Zauderer\altaffilmark{\cfa},
A.~Levan\altaffilmark{\war},
R.~Margutti\altaffilmark{\cfa},
T.~Laskar\altaffilmark{\cfa},
W.~Fong\altaffilmark{\cfa}, 
V.~Mangano\altaffilmark{\inaf}, 
D.~B.~Fox\altaffilmark{\psu},
R.~L.~Tunnicliffe\altaffilmark{\war},
R.~Chornock\altaffilmark{\cfa}, 
N.~R.~Tanvir\altaffilmark{\lei},
K.~M.~Menten\altaffilmark{\mpi},
J.~Hjorth\altaffilmark{\dcc},
K.~Roth\altaffilmark{\gem},
and T.~J.~Dupuy\altaffilmark{\cfa}
}

\altaffiltext{1}{Harvard-Smithsonian Center for Astrophysics, 60
Garden Street, Cambridge, MA 02138, USA}

\altaffiltext{2}{Department of Physics, University of Warwick,
Coventry, CV4 7AL, UK}

\altaffiltext{3}{INAF, Istituto di Astrofisica Spaziale e Fisica
Cosmica, Via U.~La Malfa 153, I-90146 Palermo, Italy}

\altaffiltext{4}{Department of Astronomy and Astrophysics, The
Pennsylvania State University, 525 Davey Lab, University Park, PA
16802, USA}

\altaffiltext{5}{Department of Physics and Astronomy, University of
Leicester, University Road, Leicester LE1 7RH, UK}

\altaffiltext{6}{Max-Planck-Institut f\"{u}r Radioastronomie, Auf dem
H\"{u}gel 69, 53121, Bonn, Germany}

\altaffiltext{7}{Dark Cosmology Centre, Niels Bohr Institute,
University of Copenhagen, Juliane Maries Vej 30, DK-2100 Copenhagen
\O, Denmark}

\altaffiltext{8}{Gemini Observatory, 670 North Aohoku Place, Hilo, HI
96720, USA}

\begin{abstract} We present the optical discovery and sub-arcsecond
optical and X-ray localization of the afterglow of the short \grb, as
well as optical, near-IR, and radio detections of its host galaxy.
X-ray observations with \swift/XRT, \chandra, and XMM-{\it Newton}
extending to $\delta t\approx 19$ d reveal a single power law decline.
The optical afterglow is faint and comparison to the X-ray flux
indicates that \grb\ is ``dark'', with a rest-frame extinction of
$A_V^{\rm host}\approx 2.5$ mag (at $z\approx 1.3$).  The intrinsic
neutral hydrogen column density inferred from the X-ray spectrum,
$N_{\rm H,int}(z=1.3)\approx 2\times 10^{22}$ cm$^{-2}$, is
commensurate with the large extinction.  The host galaxy exhibits red
optical/near-IR colors.  Equally important, JVLA observations at
$\approx 0.9-11$ d reveal a constant flux density of $F_\nu(5.8\,{\rm
GHz})=35\pm 4$ $\mu$Jy and an optically-thin spectrum, unprecedented
for GRB afterglows, but suggestive instead of emission from the host
galaxy.  The optical/near-IR and radio fluxes are well fit with the
scaled spectral energy distribution of the local ultra-luminous
infrared galaxy (ULIRG) Arp\,220 at $z\approx 1.3$, with a resulting
star formation rate of $\approx 300$ M$_\odot$ yr$^{-1}$.  The
inferred extinction and small projected offset ($2.2\pm 1.2$ kpc) are
also consistent with the ULIRG scenario, as is the presence of a
companion galaxy at a separation of about 11 kpc.  The limits on radio
afterglow emission, in conjunction with the observed X-ray and optical
emission, require a circumburst density of $n\sim 10^{-3}$ cm$^{-3}$
an isotropic-equivalent energy scale of $E_{\rm\gamma,iso}\approx
E_{\rm K,iso}\approx 7\times 10^{51}$ erg, and a jet opening angle of
$\theta_j\gtrsim 8^\circ$.  The expected fraction of luminous infrared
galaxies in the short GRB host sample is $\sim 0.01-0.3$ (for pure
stellar mass and star formation weighting, respectively).  Thus, the
observed fraction of 2 events in about $25$ hosts (GRBs 120804A and
100206A), provides additional support to our previous conclusion that
short GRBs track both stellar mass and star formation activity.
\end{abstract}


\section{Introduction}
\label{Sec:Intro}

Short-duration gamma-ray bursts (GRBs) occur in a wide range of
environments that include elliptical and star forming galaxies in the
field and in clusters (e.g., \citealt{ber11} and references therein).
These galaxies have redshifts of $z\approx 0.1$ to $\gtrsim 1$
\citep{bfp+07,rwl+10}, star formation rates of ${\rm SFR}\lesssim
0.01$ to $\approx 40$ M$_\odot$ yr$^{-1}$ \citep{ber09,pmm+11}, and
stellar masses of $M_*\approx 10^9-4\times 10^{11}$ M$_\odot$
\citep{lb10}.  These properties are suggestive of a progenitor
population that tracks both stellar mass and star formation activity
(though with a significant delay of $\sim 0.3$ Gyr; \citealt{lb10}),
in agreement with the popular compact object coalescence model (e.g.,
\citealt{elp+89,npp92}).

In a similar vein, short GRBs also exhibit a range of explosion
properties, with isotropic-equivalent energies of $E_{\rm \gamma,
iso}\sim E_{\rm K,iso}\sim 10^{49}-10^{52}$ erg
\citep{ber07,nak07,nfp09}, jet opening angles of $\theta_j\approx {\rm
few}$ to $\gtrsim 20$ deg in a few cases
\citep{bgc+06,gbp+06,sbk+06,whj+06,fbm+12}, and circumburst densities
of $n\lesssim 1$ cm$^{-3}$ \citep{bpc+05,sbk+06,fbm+12}.  The
distribution of opening angles is of particular importance since it
impacts the true energy scale and event rate.  To date, all
measurements or limits on $\theta_j$ have relied on X-ray observations
thanks to the relative brightness and high detection fraction of the
afterglow in the X-ray band compared to the optical and radio bands
\citep{nfp09,ber10}.

Here we present the optical discovery and sub-arcsecond localization
of the optical and X-ray afterglow of the short \grb, as well as
optical, near-IR, and radio detections of its host galaxy.  The
afterglow data constrain the burst properties ($E_{\rm K,iso}$,
$\theta_j$, $n$).  The host galaxy observations identify it as an
ultra-luminous infrared galaxy (ULIRG) at a photometric redshift of
$z\approx 1.3$, making \grb\ one of the most distant short bursts
known to date.  This is the first ULIRG host in the short GRB sample,
exceeding even the luminous infrared galaxy (LIRG) likely host of
GRB\,100206A \citep{pmm+11}.  We present the afterglow and host galaxy
observations in \S\ref{sec:obs}, extract the explosion properties in
\S\ref{sec:ag}, and determine the host galaxy photometric redshift and
properties in \S\ref{sec:host}.  Throughout the paper we report
magnitudes in the AB system (unless otherwise noted), use a Galactic
extinction value of $E(B-V)\approx 0.204$ mag \citep{sf11}, and employ
the standard cosmological parameters: $H_{0}=71$ km s$^{-1}$
Mpc$^{-1}$, $\Omega_{\Lambda}= 0.73$, and $\Omega_{\rm M}=0.27$.

\section{Observations and Analysis}
\label{sec:obs}

\grb\ was discovered with the \swift\ Burst Alert Telescope (BAT) on
2012 August 4 at 00:55:47.8 UT \citep{GCN13581}, and was also detected
with Konus-WIND \citep{GCN13614}.  The burst duration is $T_{\rm 90}=
0.81\pm 0.08$ s ($15-350$ keV) with a fluence of $F_\gamma=(8.8\pm
0.5)\times 10^{-7}$ erg cm$^{-2}$ ($15-150$ keV) and $(1.45\pm
0.30)\times 10^{-6}$ erg cm$^{-2}$ ($15-1000$ keV).  A joint analysis
of the BAT and Konus-WIND data indicates a peak energy of
$E_p=135^{+66}_{-29}$ keV \citep{GCN13614}.  The spectral lags are
$16\pm 12$ ms ($15-25$ to $50-100$ keV) and $-5\pm 6$ ms ($25-50$ to
$100-350$ keV), typical of short GRBs \citep{GCN13585}.

\swift/X-ray Telescope (XRT) observations commenced about 78 s after
the burst and led to the identification of a fading source, located at
RA=\ra{15}{35}{47.51}, Dec=\dec{$-$28}{46}{56.9} with an uncertainty
of 1.4'' radius (90\% containment, UVOT-enhanced; \citealt{gcn13577}).
Observations with the UV/Optical Telescope (UVOT) began about 97 s
after the burst, but no counterpart was detected to a $3\sigma$ limit
of $\gtrsim 21.4$ mag in the white filter (at $\delta t\approx 97-247$
s; \citealt{GCN13582}).  Optical and near-IR observations with GROND
starting at $\delta t\approx 1.5$ hr also led to non-detections with
$\gtrsim 22$ mag ($griz$) and $\gtrsim 20.6$ mag ($J$-band;
\citealt{GCN13579}).

\subsection{X-ray Observations}
\label{sec:xray}

We analyze the \swift/XRT data using the HEASOFT package (v6.11) and
latest calibration files with the standard filtering and screening
criteria.  We generate the $0.3-10$ keV count-rate light curve
following the procedure described in \citet{mzb+12}, with a re-binning
scheme that ensures a minimum signal-to-noise ratio of $4$ for each
temporal bin.  The data comprise 33 s ($\delta t\approx 97-130$ s) in
Windowed Timing (WT) mode, and 18 ks in Photon Counting (PC) mode
($\delta t\approx 150-6\times 10^4$ s).

We fit the time-averaged WT spectrum with an absorbed power-law model
($tbabs\times ztbabs\times pow$ in \emph{Xspec}) using a Galactic
neutral hydrogen column density of $N_{\rm H,MW}\approx 9.3\times
10^{20}$ cm$^{-2}$ \citep{kbh+05}.  The resulting spectral photon
index is $\Gamma=(2.5\pm 0.3)$ and the excess neutral hydrogen column
density is $N_{\rm H,int}=(3.5\pm 1.1)\times 10^{21}$ cm$^{-2}$ at
$z=0$ (${\rm C-stat}=92$ for 122 degrees of freedom; uncertainties are
$1\sigma$).  From the time-averaged PC spectrum we infer $\Gamma=
(2.1\pm 0.1)$ and $N_{\rm H,int}=(3.2\pm 0.5)\times 10^{21}$ cm$^{-2}$
at $z=0$ (${\rm C-stat}=283$ for 316 degrees of freedom).  We adopt
the latter value of $N_{\rm H,int}$ in the time-resolved spectral
analysis, designed to account for the source spectral evolution and
the resulting count-to-flux conversion factor \citep{mzb+12}.  The
uncertainties arising from the flux calibration are properly
propagated in the unabsorbed $0.3-10$ keV flux light curve.  We note
that at $z\approx 1.3$ (see \S\ref{sec:host}) the best-fit parameters
are $\Gamma\approx 1.9$ and the intrinsic neutral hydrogen column
density is $N_{\rm H,int}\approx 2\times 10^{22}$ cm$^{-2}$.

We also analyze a \chandra\ ACIS-S observation obtained on 2012 August
13.45 UT ($\delta t\approx 9.41$ d) with a total exposure time of 19.8
ks (PI: Troja; \citealt{GCN13640}).  The X-ray afterglow is detected
with a significance of about $13\sigma$ at a count rate of $1.6\pm
0.3$ cps in the $0.5-8$ keV range.  Adopting the best-fit spectral
parameters from the XRT analysis this translates to an unabsorbed
$0.3-10$ keV flux of $(2.8\pm 0.5)\times 10^{-14}$ erg s$^{-1}$
cm$^{-2}$.

We further obtained an XMM-{\it Newton} observation on 2012 August
22.91 UT ($\delta t\approx 18.9$ d; PI: Margutti) to search for the
signature of a jet break at late time.  We analyze the EPIC data with
the XMM Science Analysis System (SAS v11.0.0), selecting events with
${\rm PATTERN}\le 12$ for the MOS cameras, ${\rm PATTERN}\le 4$ for
the pn camera, and ${\rm FLAG}=0$ for both.  To reduce the
contamination by soft proton flares, we screen the original event
files using a sigma-clipping algorithm.  The remaining good science
time is 29.5 ks for MOS1 and MOS2, and 25.5 ks for pn.  The X-ray
afterglow is detected in the MOS1 and MOS2 images with $(1.0\pm
0.3)\times 10^{-3}$ cps ($0.2-10$ keV) and in the pn image at the with
$(4.3\pm 0.7)\times 10^{-3}$ cps ($0.2-10$ keV).  We perform spectral
analysis using the {\tt evselect} tool, with the response files
generated with the {\tt rmfgen} and {\tt arfgen} tools.  The resulting
inter-calibration factor of MOS1 and MOS2 relative to pn is $0.9-1.1$.
We adopt the value of $N_{\rm H,int}$ from the XRT analysis, leading
to a pn unabsorbed $0.3-10$ keV flux of $(2.4\pm 0.5)\times 10^{-14}$
erg s$^{-1}$ cm$^{-2}$.

\subsection{Optical Afterglow Discovery and Relative X-ray Astrometry}
\label{sec:optag}

\begin{figure} 
\centering
\includegraphics[angle=0,height=2.3in]{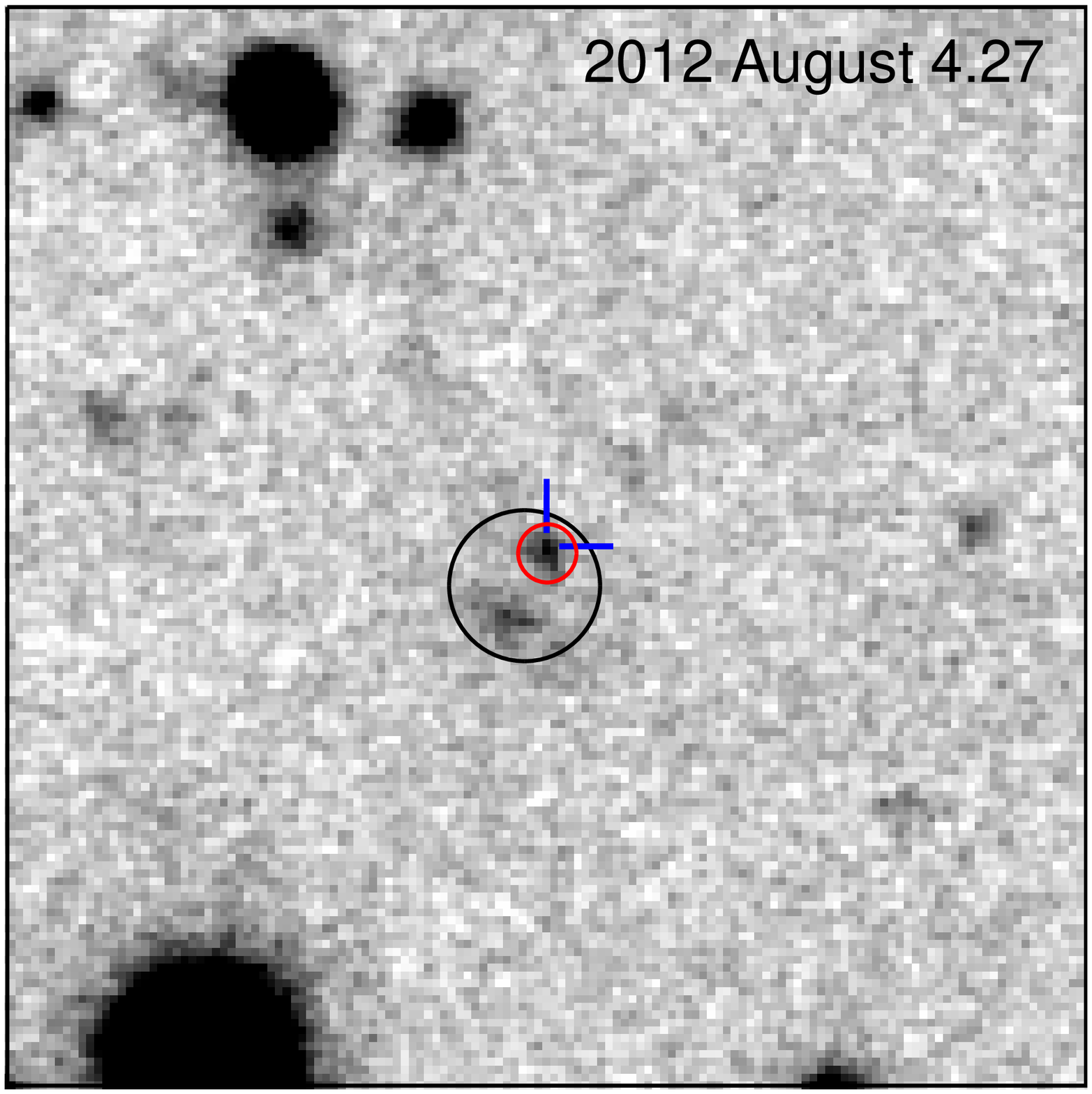}\hfill
\includegraphics[angle=0,height=2.3in]{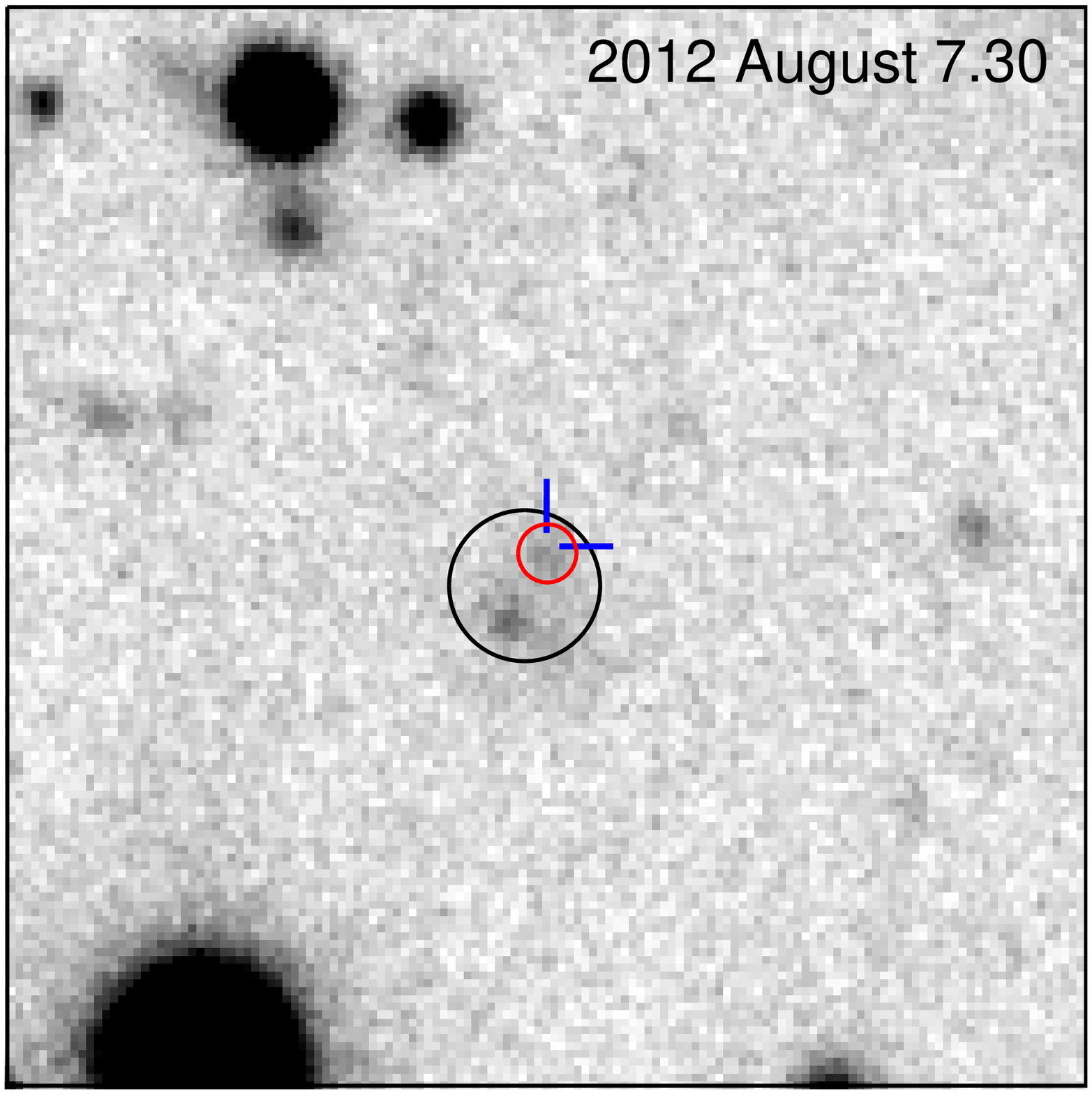}\hfill
\includegraphics[angle=0,height=2.3in]{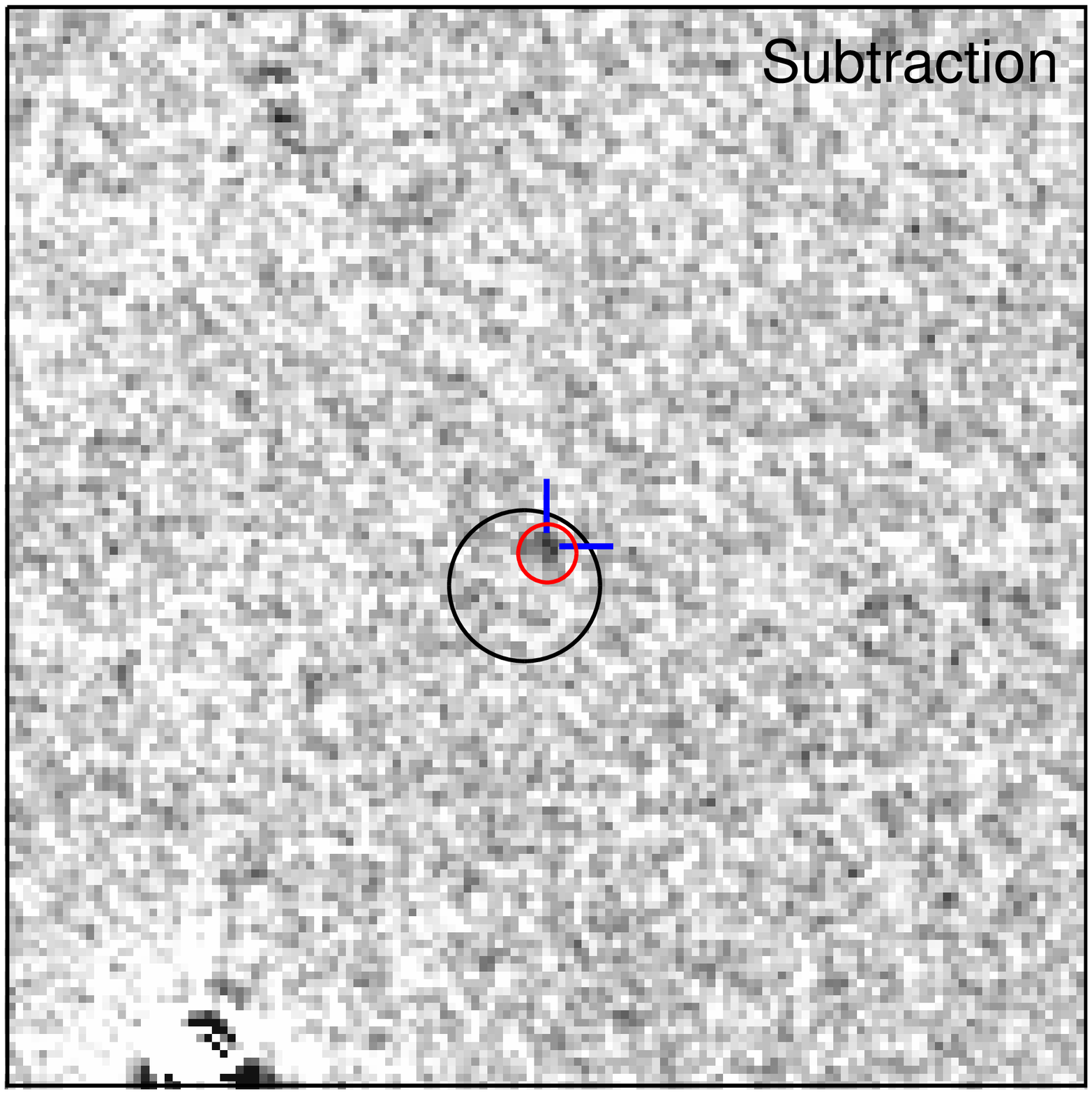}
\caption{Gemini-North $i$-band images centered on the location of the
\swift/XRT error circle (black circle; 1.4'' radius 90\% containment)
obtained at $\delta t\approx 0.23$ d (left) and $\delta t\approx 3.26$
d (center), along with a subtraction of the two epochs (right).  The
subtraction reveals a fading optical afterglow (cross-hairs).  The
\chandra\ X-ray afterglow position coincides with the optical position
(red circle; $0.54''$ radius 95\% containment).  Also seen is a galaxy
located at a projected angular distance of $\approx 1.4''$ south-east
of the host galaxy position.  Each image is 20'' on a side.
\label{fig:image}}
\end{figure}

We obtained two epochs of $i$-band imaging with the Gemini
Multi-Object Spectrograph (GMOS; \citealt{hja+04}) on the Gemini-North
8-m telescope on 2012 August 4.27 and 7.30 UT ($\delta t\approx 0.23$
d and $\approx 3.26$ d, respectively).  The observations consisted of
1980 s and 2880 s, respectively, in 0.65'' seeing.  We process the
data using the {\tt gemini} package in IRAF, and perform photometry
using a zero-point of $28.46\pm 0.10$ mag measured on the nights of
August $2-8$ UT.  We perform digital image subtraction of the two
epochs with the ISIS package \citep{ala00}, and recover a fading
source with $m_i=26.2\pm 0.2$ AB mag (with an additional systematic
uncertainty of $\pm 0.1$ mag due to the zero-point uncertainty).  We
consider this source to be the optical afterglow of \grb.  Corrected
for Galactic extinction the resulting flux density is $F_\nu=0.17\pm
0.05$ $\mu$Jy.  Images of the two epochs and the resulting subtraction
are shown in Figure~\ref{fig:image}.

We determine the absolute position of the afterglow by astrometrically
matching the images to the 2MASS reference frame using 45 common
sources.  The resulting astrometric uncertainty is 0.15'' in each
coordinate.  The afterglow position in the residual image is
RA=\ra{15}{35}{47.479}, Dec=\dec{$-$28}{46}{56.17} (J2000), with a
centroid uncertainty of about 0.05'' in each coordinate.

To locate the \chandra\ X-ray afterglow on the optical images we
perform differential astrometry using 4 common sources.  We find a
relative offset between the two coordinate frames (\chandra\ to
Gemini) of $\delta{\rm RA}=-0.06\pm 0.21''$ and $\delta{\rm Dec}=
-0.01\pm 0.17''$, leading to a refined X-ray afterglow position of
RA=\ra{15}{35}{47.478}, Dec=\dec{$-$28}{46}{56.30} with an uncertainty
of about 0.35'' radius that takes into account a centroid uncertainty
of about 0.06''; see Figure~\ref{fig:image}).  The \chandra\ position
is in excellent agreement with the optical afterglow position.

\subsection{Optical/Near-IR Host Galaxy Observations} 
\label{sec:oir}

In the second Gemini observation we detect an extended source near the
afterglow position with $m_i=24.80\pm 0.15$ mag (including zero-point
uncertainty and corrected for Galactic extinction) located at
RA=\ra{15}{35}{47.477}, Dec=\dec{$-$28}{46}{56.44}, with a centroid
uncertainty of about 0.10''.  We consider this source to be the host
galaxy of \grb.  Given the brightness of the galaxy, the probability
of chance coincidence using a radius of $1''$ (e.g.,
\citealt{bkd02,ber10}) is $P_{\rm cc}\approx 0.02$.  We investigate
potential association with brighter galaxies ($\approx 19-21$ mag) at
larger offsets ($\approx 0.2-1'$) but find chance coincidence
probabilities of $P_{\rm cc}\approx 0.2-0.7$, indicating that these
are not likely to be associated with \grb.

We also observed \grb\ with GMOS on the Gemini-North telescope on 2012
August 4.24 UT ($\delta t\approx 0.20$ d) in $r$-band with a total
exposure time of 2340 s.  We process the data using the {\tt gemini}
package in IRAF, and perform photometry using a zero-point of
$28.41\pm 0.02$ mag measured on the nights of July 27 -- August 9 UT.
Photometry at the position of the host galaxy reveals a faint source
with $m_r=26.2\pm 0.2$ mag, corresponding to a flux density of
$F_\nu=0.19\pm 0.04$ $\mu$Jy (corrected for Galactic extinction).
Since this was an early observation, the measurement may be
contaminated by afterglow emission.  However, taking into account a
spectral shape of $F_\nu\propto\nu^{-0.6}$ and the large rest-frame
extinction (see \S\ref{sec:ag}) we conclude that the afterglow
contribution is sub-dominant, $\approx 0.03$ $\mu$Jy.

We observed the host galaxy in $J$-band with the FourStar near-IR
camera on the Magellan/Baade 6.5-m telescope on 2012 August 28.98 UT
with a total on-source time of 2390 s.  We analyze the data using a
custom pipeline in python, and perform photometry using common sources
with the 2MASS catalog.  The host galaxy has a measured brightness of
$m_J=23.05\pm 0.20$ mag (corrected for Galactic extinction).

We obtained $Y$- and $K_s$-band observations with the High Acuity
Wide-field K-band Imager (HAWK-I) on the VLT starting on 2012
September 7.96 UT, with a total exposure time of 1320 s in each
filter.  We produce dark-subtracted and flat-fielded images of the
field using the HAWK-I pipeline within {\tt esorex}, and perform
photometry on the $K_s$-band image relative to the 2MASS catalog.  We
determine the $Y$-band calibration using the instrumental zero-point
(appropriate for our observations, which were obtained in good
conditions), and confirm that this is appropriate by extrapolating
2MASS $J$-band photometry to observations taken in the $i$-band.  The
host galaxy is detected in the $K_s$-band with $m_K=22.0\pm 0.1$ mag,
and weakly in $Y$-band with $m_Y=23.7\pm 0.3$ mag (both values are
corrected for Galactic extinction).

Finally, we obtained spectroscopic observations of the host galaxy
with the FOcal Reducer and low dispersion Spectrograph (FORS2) on the
VLT on 2012 August 19.01 UT.  The observations consisted of $4\times
600$ s exposures, covering the wavelength range $4300-9300$ \AA.  We
set the slit position angle to $149^\circ$ to cover the host and
nearby galaxy (see Figure~\ref{fig:image}).  We detect no continuum or
line emission at the position of the host galaxy, and only a faint
continuum from the nearby galaxy.

The host galaxy photometry is summarized in Table~\ref{tab:host}.  We
also provide photometry for the extended source located $\approx
1.4''$ to the south-east of the host galaxy position
(Figure~\ref{fig:image}).

\subsection{Radio Observations} 
\label{sec:radio}

We observed \grb\ with the Karl G.~Jansky Very Large Array (JVLA)
starting on 2012 August 4.97 UT ($\delta t\approx 0.93$ d) at a mean
frequency of 5.8 GHz.  We utilized the WIDAR correlator \citep{pcb+11}
with a bandwidth of about 1 GHz in each sideband, centered at 4.9 and
6.7 GHz.  All observations were undertaken in the B configuration,
utilizing 3C286 for bandpass and flux calibration, with interleaved
observations of J1522$-$2730 for gain calibration.  We calibrate and
analyze the data using standard procedures in the Astronomical Image
Processing System (AIPS; \citealt{gre03}), and list the resulting flux
density measurements in Table~\ref{tab:evla}.

In the first four epochs ($\delta t\approx 0.9-10.9$ d) we detect a
single unresolved source coincident with the optical and X-ray
afterglow, as well as the host galaxy: RA=\ra{15}{35}{47.485} ($\pm
0.008$), Dec=\dec{$-$28}{46}{56.44} ($\pm 0.20$).  However, the source
remains constant in brightness, with flux densities of
$F_\nu(4.9\,{\rm GHz})= 43\pm 4$ $\mu$Jy and $F_\nu(6.7\,{\rm
GHz})=25\pm 4$ $\mu$Jy.  The steady brightness and optically thin
spectrum, $F_\nu\propto \nu^{-1.7\pm 0.8}$, are unprecedented for GRB
radio afterglows at early time \citep{gs02}, but are instead
suggestive of emission from the host galaxy.  Indeed, the centroid of
the radio source shows a smaller offset relative to the optical host
galaxy centroid ($0.10''$) than to the optical afterglow position
($0.28''$).  The probability of chance coincidence for a source of
this brightness within $\sim 1''$ of the optical/near-IR host galaxy
position is $P_{\rm cc}\approx 2\times 10^{-4}$, based on the number
counts of faint 5 GHz sources \citep{fwk+91}.  This indicates that the
radio source is the host galaxy of \grb.  The observation on 2012
September 11 UT was obtained during a JVLA re-configuration leading to
poorer noise characteristics in the resulting image.  The $3\sigma$
limits are consistent with the earlier detections of the steady
source.

We use the lack of variability to place a $3\sigma$ upper limit on the
radio afterglow brightness of $F_\nu(5.8\,{\rm GHz})\lesssim 20$
$\mu$Jy.

In addition, we obtained JVLA observations of the LIRG host galaxy of
GRB\,100206A \citep{pmm+11} on 2012 September 10.27 UT to determine
whether its large total star formation rate inferred from optical/IR
data ($\sim 40$ M$_\odot$ yr$^{-1}$) produces radio emission.  The
observing setup and data analysis follow the procedure described
above.  We observed 3C48 for flux and bandpass calibration, and
interleaved observations of J0238$+$1636 for gain calibration.  There
is a bright, contaminating source in the field (27 mJy at 4.9 GHz and
7 mJy at 6.7 GHz), located about $4.5'$ from the position of
GRB\,100206A.  We therefore image the field utilizing self-calibration
techniques on this bright source, leading to a non-detection of radio
emission with a conservative $5\sigma$ limit of $F_\nu(5.8\,{\rm
GHz})\lesssim 0.1$ mJy

\section{Afterglow Properties}
\label{sec:ag}

We model the X-ray data, optical detection, and radio upper limits
using the standard afterglow synchrotron model \citep{gs02}.  We
follow the standard assumptions of synchrotron emission from a
power-law distribution of electrons ($N(\gamma)\propto\gamma^{-p}$ for
$\gamma\ge\gamma_m$) with constant fractions of the post-shock energy
density imparted to the electrons ($\epsilon_e$) and magnetic fields
($\epsilon_B$).  The additional free parameters of the model are the
isotropic-equivalent blast-wave kinetic energy ($E_{\rm K,iso}$), and
the circumburst density ($n$ for a constant density medium).

We note three key observational facts to guide the afterglow modeling.
First, the X-ray light curve is best fit with a single decline rate of
$\alpha_X=-0.93\pm 0.06$, which coupled with the spectral index of
$\beta_X\approx -0.9\pm 0.1$ (\S\ref{sec:xray}) indicates that the
synchrotron cooling frequency is $\nu_c\sim \nu_X$ and that $p\approx
2.2$.  Second, the peak of the X-ray afterglow is $\approx 10-20$
$\mu$Jy at $\delta t\approx 200$ s.  This suggests that the radio
light curve will eventually reach a similar peak flux density if the
synchrotron peak frequency is $\nu_m\approx\nu_X$ at $\delta t\approx
200$ s.  Such a peak flux density is consistent with our inferred
radio upper limits.  Finally, the optical and X-ray flux densities at
$\delta t\approx 5.5$ hr are comparable, indicating that the optical
to X-ray spectral index is $\beta_{\rm OX}\approx 0$, compared to an
expected slope of $\approx 0.6$ (for $p=2.2$ and $\nu_c\sim \nu_X$).
This shallow slope indicates that \grb\ can be classified as a
``dark'' burst, following the definition of \citet{jhf+04}.

\begin{figure}
\centering
\includegraphics[angle=0,width=6in]{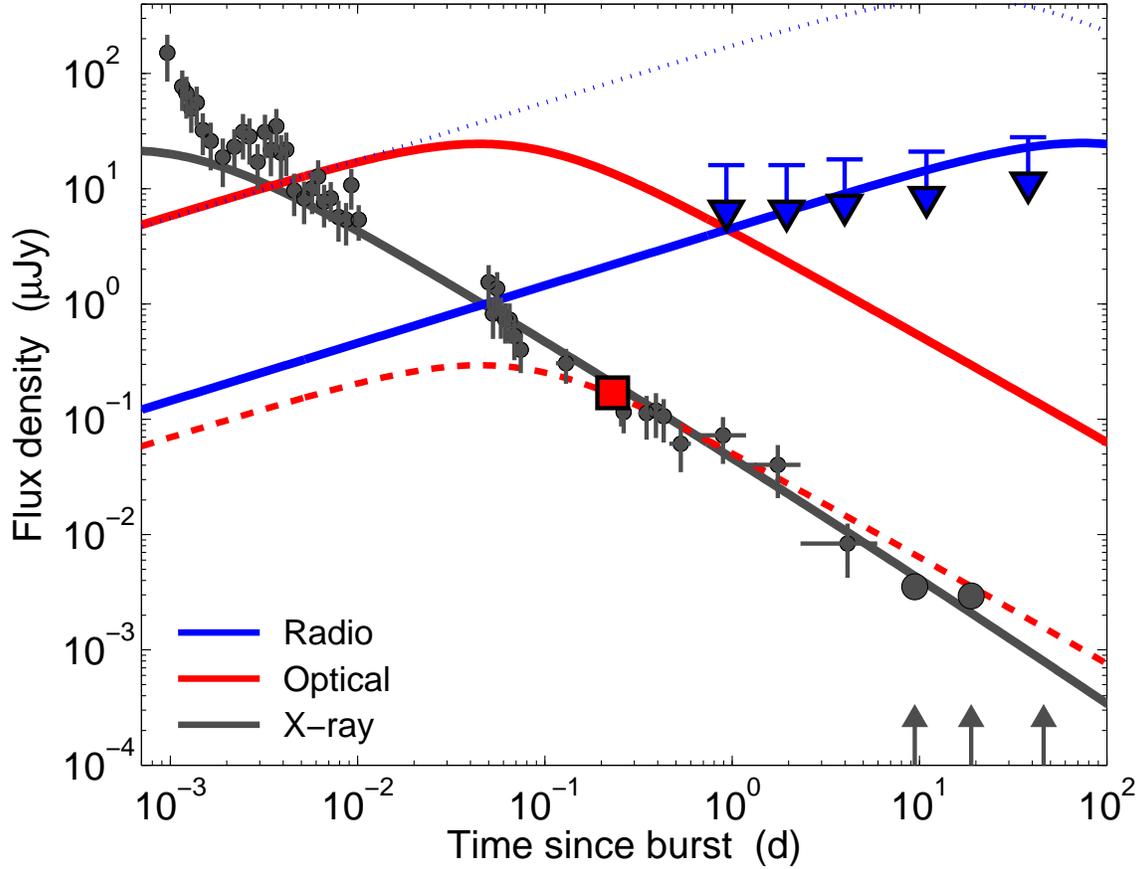}
\caption{X-ray light curve, optical $i$-band detection, and radio 5.8
GHz upper limits for the afterglow of \grb.  The solid lines are an
afterglow model fit using the formulation of \citet{gs02}, with
$E_{\rm K,iso}\approx 8\times 10^{51}$ erg, $n\approx 10^{-3}$
cm$^{-3}$, $\epsilon_e\approx 0.3$, $\epsilon_B\approx 0.1$, and
$p\approx 2.2$.  The low observed flux density in the optical band is
suggestive of extinction, and can be explained with $A_{V}^{\rm
host}\approx 2$ mag (dashed red line).  The dotted blue line indicates
the predicted radio light curve for a model with fixed parameters of
$n=1$ cm$^{-3}$ and $E_{\rm K,iso}=6\times 10^{51}$ erg (i.e., matched
to $E_{\rm \gamma,iso}$), that produces an indistinguishable fit in
the X-rays (this model requires $\epsilon_e\approx 0.25$,
$\epsilon_B\approx 0.1$, and $p\approx 2.1$).  Clearly, such a high
density can be ruled out.  The arrows mark the \chandra\ and XMM-{\it
Newton} observations, including the time of a late \chandra\
observation.
\label{fig:ag}}
\end{figure}

Guided by these results, we find that the data can be fit with the
following parameters (using $z=1.3$; see \S\ref{sec:host}): $E_{\rm
K,iso}\approx 8\times 10^{51}$ erg, $n\approx 10^{-3}$ cm$^{-3}$,
$\epsilon_e\approx 0.3$, $\epsilon_B\approx 0.1$, and $p\approx 2.2$.
The inferred blast-wave energy is comparable to the
isotropic-equivalent $\gamma$-ray energy, $E_{\rm \gamma,iso}\approx
6\times 10^{51}$ erg.  To explain the suppressed optical emission we
also require $A_V^{\rm host}\approx 2.5$ mag (at $z=1.3$).  The
resulting light curves are shown in Figure~\ref{fig:ag}.  Models with
higher values of $E_{\rm K,iso}$ and/or $n$ lead to a lower value of
$\nu_m$ and a larger peak flux density, and therefore violate the
radio limits (see for example dotted line in Figure~\ref{fig:ag}).

In addition, the lack of a break in the X-ray light curve to at least
$\approx 20$ d places a lower bound on the jet collimation angle, with
$\theta_j\gtrsim 8^\circ$, where we use the values of $E_{\rm K,iso}$
and $n$ inferred above.  This indicates a beaming correction factor of
$f_b^{-1}\equiv [1-{\rm cos}(\theta_j)]^{-1}\lesssim 120$, and hence a
beaming-corrected energy scale of $E_\gamma+E_K\approx 1.1\times
10^{50}-1.4\times 10^{52}$ erg; the upper bound is set by isotropy.

Since in the afterglow model above, the X-ray light curve sets the
overall flux density scale, an alternative explanation for the radio
non-detections and the low optical flux density is that the X-ray
emission is dominated by a different emission mechanism.  One
possibility is contribution from inverse Compton emission
\citep{se01}, but this requires a large density of $n\gtrsim 10^{2}$
cm$^{-3}$.  Another possibility is emission from a newly-born
rapidly-spinning magnetar (e.g., \citealt{zm01}), but the expected
evolution in this case is a relatively constant brightness for a
duration similar to the spin-down timescale, followed by $F_\nu\propto
t^{-2}$ at later time (for a typical braking index of $3$).  This is
quite distinct from the observed single power law of $F_\nu\propto
t^{-1}$ at $\delta t\gtrsim 200$ s.

\section{A ULIRG Host Galaxy}
\label{sec:host}

The observed optical/near-IR spectral energy distribution (SED) of the
host galaxy exhibits a red color of $i-K\approx 2.8$ mag ($\approx
4.3$ Vega mag); the less certain $r-K$ color is $\approx 3.9$ mag
($\approx 5.6$ Vega mag).  There is also noticable steepening between
the $i$- and $Y$-band filters.  These properties are indicative of a
Balmer/4000 \AA\ break at $z\approx 1.3$ (with a range of $\approx
1.1-1.5$) and either an old or dusty stellar population.

\begin{figure}
\centering
\includegraphics[angle=0,width=3.5in]{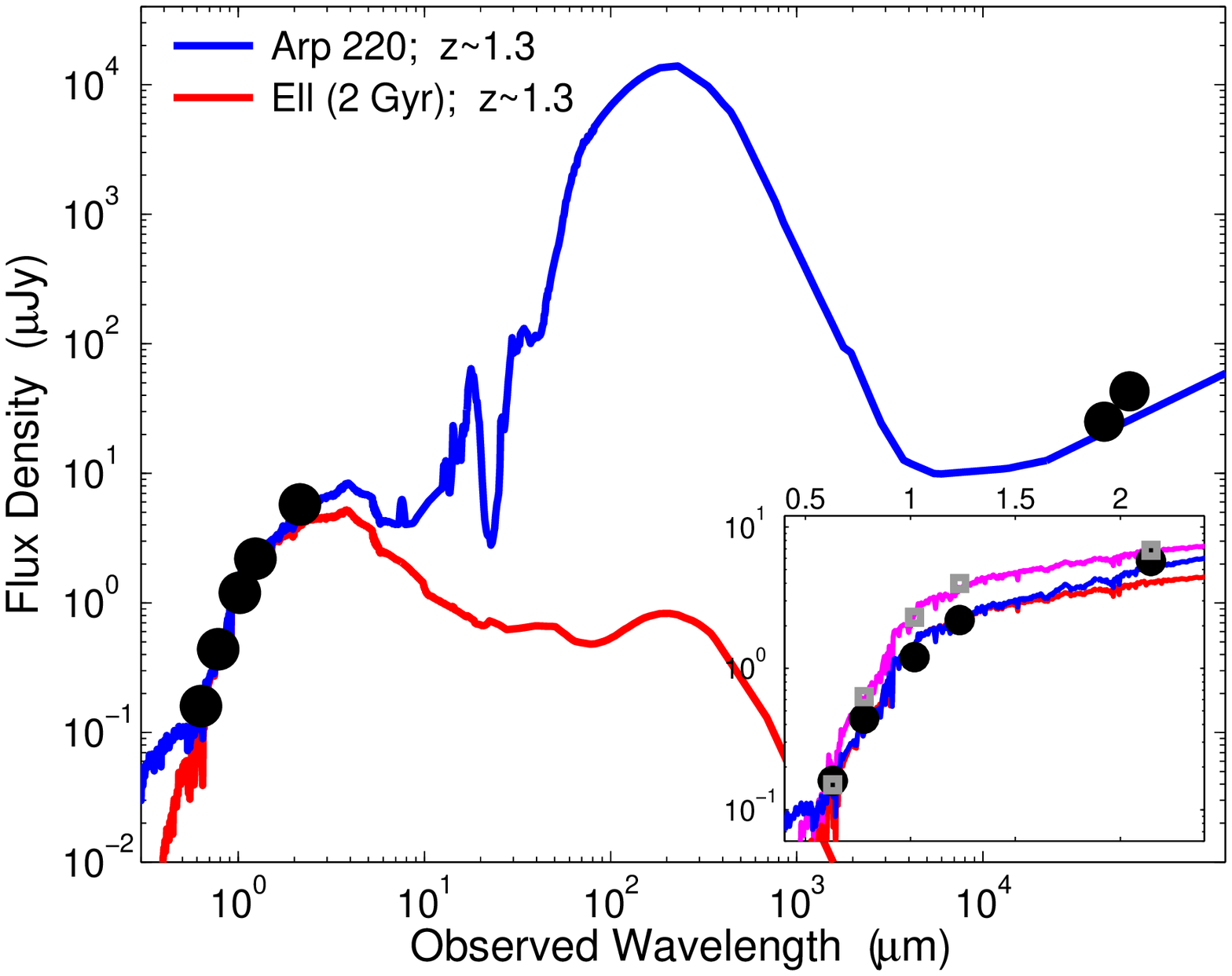}\hfill
\includegraphics[angle=0,width=3.5in]{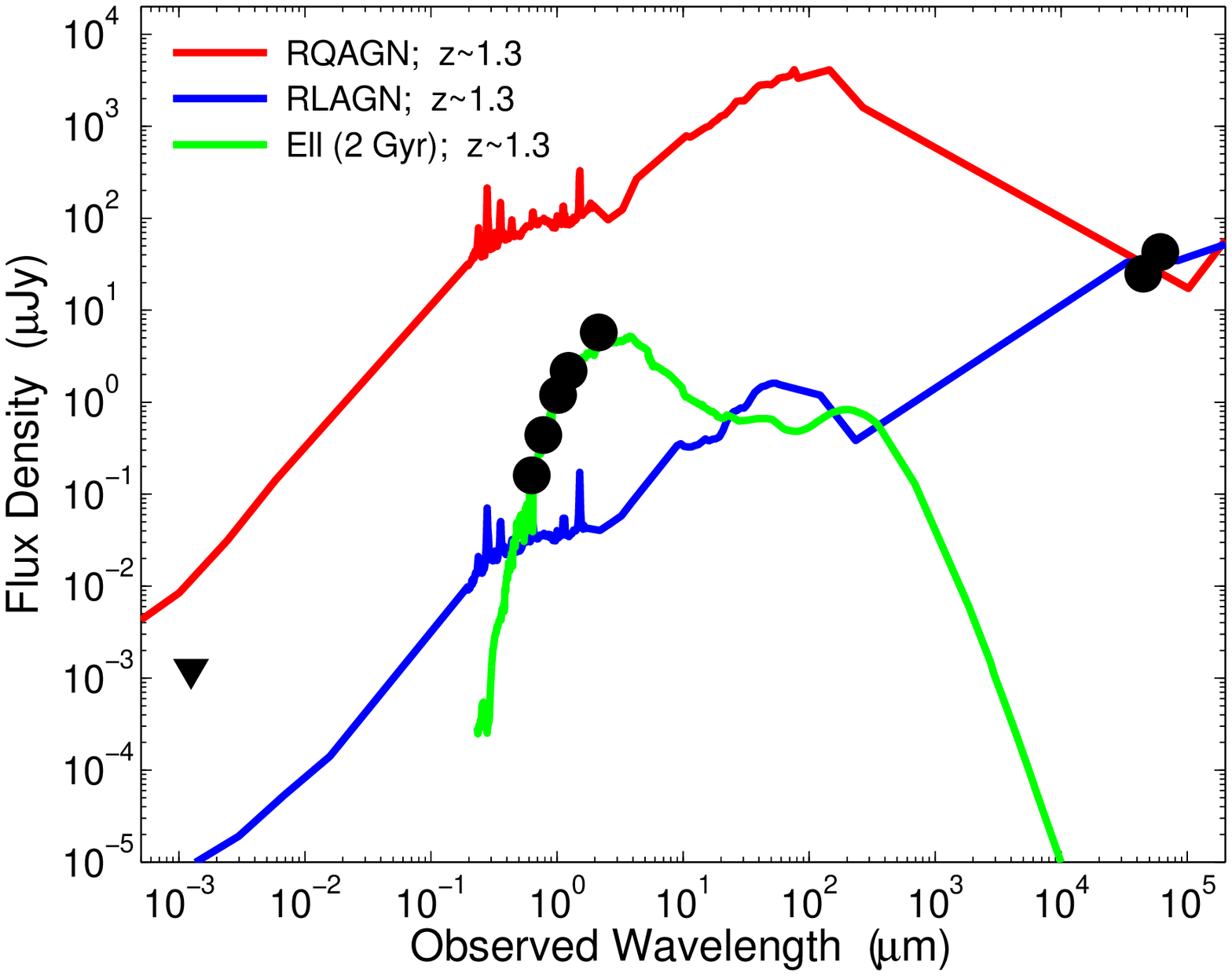}
\caption{{\it Left:} Spectral energy distribution of the host galaxy
of \grb\ (black circles: optical, near-IR, and radio) compared to a
scaled SED of Arp\,220 (blue line) and an elliptical galaxy with a 2
Gyr stellar population (red line) at $z=1.3$.  Both SEDs are scaled to
match the optical/near-IR photometry of the host galaxy.  The inset
shows a zoom-in on the optical/near-IR range, highlighting that both
SEDs provide a reasonable fit in this wavelength range (with a
somewhat better fit in $K_s$-band for the Arp\,220 SED).  The gray
symbols mark the photometry of the nearby galaxy, along with an
elliptical galaxy model at $z=1.3$.  In both cases, the apparent
steepening between the $i$- and $Y$-band filters points to a similar
redshift.  {\it Right:} Comparison to the scaled SEDs of radio-quiet
(red line) and radio-loud (red line) AGN at $z=1.3$, as well as the
SED of an elliptical galaxy with a 2 Gyr stellar population (green
line).  The AGN SEDs are scaled to match the observed radio flux
density of the host galaxy.  The radio-quiet AGN scenario
over-predicts the optical, near-IR, and X-ray fluxes.  On the other
hand, the radio-loud scenario matches the radio flux density without
violating the other measurements; the optical/near-IR emission is
dominated by stellar emission.
\label{fig:sed}}
\end{figure}

Taken in conjunction with the radio detection, the SED is reminiscent
of ULIRGs ($L_{\rm FIR}\gtrsim 10^{12}$ L$_\odot$).  To investigate
this possibility we compare the host galaxy fluxes to the
SED\footnotemark\footnotetext{Obtained from the SWIRE Template Library
\citep{ptm+07}.  We note that ULIRG templates exhibit some
variability, but this will have little effect on the resulting
photometric redshift.} of the local ULIRG Arp\,220.  As shown in
Figure~\ref{fig:sed}, at $z\approx 1.3$ a simple scaling of Arp\,220
provides a remarkable fit to the data.  An elliptical galaxy template
(2 Gyr old population; \citealt{ptm+07}) at $z\approx 1.3$ provides a
reasonable fit in the optical/near-IR (although it underestimates the
observed $r$- and $K_s$-band fluxes), but cannot explain the radio
emission (Figure~\ref{fig:sed}).

At $z\approx 1.3$ the rest-frame $B$-band absolute magnitude is
$M_B\approx -20.2$ mag, or $L_B\approx 0.2\,L^*$ in comparison to the
$B$-band luminosity function at $z\sim 1.3-2$ \citep{itz+05}.  The
rest-frame $K$-band absolute magnitude is $M_K\approx -22.1$ mag
($-24.0$ Vega mag, or $L_K\approx 0.4\,L^*$; \citealt{cmd+06}).  This
luminosity corresponds to a stellar mass of $M_*5\times \approx
10^{10}$ M$_\odot$ (using a characteristic mass-to-light ratio of
$M_*/L_K\approx 0.3$; \citealt{dbf+04}).  The infrared bolometric
luminosity scaled using the SED of Arp\,220 is $L_{\rm FIR}\approx
10^{12}$ L$_\odot$.

The unobscured star formation rate is inferred from the observed
$r$-band ($\lambda_0\approx 2700$ \AA) to be $\approx 1$ M$_\odot$
yr$^{-1}$.  However, from the observed radio emission we estimate the
total star formation rate to be much larger \citep{yc02}:
\begin{equation}
{\rm SFR}\approx \frac{F_{\rm\nu,\mu Jy}\,d_{\rm L,Gpc}^2/(1+z)}
{25\nu_{\rm 0,GHz}^{-0.75}+0.7\nu_{\rm 0,GHz}^{-0.1}}\approx 
300\,\,\,\,{\rm M}_\odot\,\,{\rm yr}^{-1},
\label{eqn:sfr}
\end{equation} 
where $\nu_0=(1+z)\nu_{\rm obs}$ is the rest frequency, and $d_L$ is
the luminosity distance.  Assuming that the host stellar mass has been
assembled with this star formation rate gives a characteristic age of
about $0.15$ Gyr.  Using the limit on radio emission from GRB\,100206A
in Equation~\ref{eqn:sfr} we find ${\rm SFR}\lesssim 60$ M$_\odot$
yr$^{-1}$.

The large star formation rate in the host of \grb\ is also expected to
produce X-ray emission, with $L_X\approx 7\times 10^{39}\,\times\,{\rm
SFR}$ erg s$^{-1}$ \citep{whj+04,vtm+12}.  For the values inferred
above we find an expected luminosity of $L_X\approx 2\times 10^{42}$
erg s$^{-1}$, corresponding to a flux of $F_X\approx 4.5\times
10^{-16}$ erg s$^{-1}$ cm$^{-2}$.  This is about 55 times lower than
the afterglow flux measured with XMM-{\it Newton} at $\delta t\approx
19$ d, and hence consistent with a star formation origin for the radio
emission.

An alternative interpretation of the radio detection is emission from
an active galactic nucleus (AGN).  Matching the standard template of a
radio-quiet AGN \citep{sbw+11} to the observed radio flux density
(using $z=1.3$), leads to a substantial over-estimate of the
optical/near-IR brightness, and a much bluer color
(Figure~\ref{fig:sed}).  Similarly, the expected X-ray flux in this
scenario is $F_X\approx 6.4\times 10^{-14}$ erg cm$^{-2}$ s$^{-1}$,
several times brighter than the XMM-{\it Newton} detection of the
afterglow.  Thus, we can rule out a radio-quiet AGN origin.  If we
instead use the standard radio-loud AGN template \citep{sbw+11}
matched to the radio flux density, we find that the expected
optical/near-IR and X-ray fluxes are $\sim 1-2$ orders of magnitude
fainter than measured (Figure~\ref{fig:sed}).  Indeed, the observed
lower bound on the ratio of radio to X-ray luminosity, $\nu L_{\rm\nu,
rad}/L_X\gtrsim 2\times 10^{-4}$, along with an upper bound of
$L_X\lesssim 5\times 10^{44}$ erg s$^{-1}$ are in good agreement with
samples of low-luminosity AGN (e.g., \citealt{tw03}).  In this
scenario, the optical/near-IR emission is instead dominated by a
stellar component, either from an old population or a reddened young
population with a modest star formation rate of a few M$_\odot$
yr$^{-1}$.

While both a ULIRG and an AGN origin can explain the observed radio
emission, we note that the former also offers a natural explanation
for the inferred extinction, and fits the broad-band host SED with a
single component.  We therefore consider a ULIRG as the more likely
explanation for the host galaxy of \grb.

The offset between the optical afterglow position and host galaxy
centroid is $0.27\pm 0.15''$, corresponding to $2.2\pm 1.2$ kpc at
$z\approx 1.3$.  This is a relatively small offset for short GRBs
(with a median of $\approx 5$ kpc; \citealt{fbf10,ber10}), though not
unprecedented.  Still, in the context of a ULIRG origin, the small
offset is consistent with the inferred afterglow extinction.

Finally, we determine a photometric redshift for the galaxy located
$\approx 1.4''$ from the host galaxy position and find $z\approx 1.3$,
consistent with that of the host galaxy (Figure~\ref{fig:sed}).  Thus,
it appears likely that these galaxies, with a projected separation of
about 11 kpc, are interacting or merging.  This is not unexpected
since at least some ULIRG activity is triggered by galaxy mergers.

\section{Summary and Conclusions}
\label{Sec:Conc}

We presented the discovery and sub-arcsecond localization of the
optical and X-ray afterglow of the short \grb.  A comparison of the
observed fluxes points to substantial rest-frame extinction of
$A_V^{\rm host}\approx 2.5$ mag, commensurate with the large neutral
hydrogen column density, $N_{\rm H,int}\approx 2\times 10^{22}$
cm$^{-3}$.  In conjunction with deep radio limits we infer an energy
of $E_{\rm\gamma,iso}\approx E_{\rm K,iso}\approx 7\times 10^{51}$ erg
and a low circumburst density of $n\sim 10^{-3}$ cm$^{-3}$.  The lack
of a break in the X-ray afterglow at $\lesssim 20$ d, leads to
$\theta_j\gtrsim 8^\circ$, in line with existing measurements of short
GRB jets \citep{fbm+12}.  We note that a 60 ks \chandra\ observation
obtained at $\delta t\approx 46$ d (PI: Burrows) may extend this limit
to $\gtrsim 11^\circ$ if no break is detected (i.e., an expected flux
of $\approx 7\times 10^{-15}$ erg s$^{-1}$ cm$^{-2}$), or determine an
angle of $\theta_j\approx 10^\circ$ if a break at an intermediate time
of $\approx 33$ d is detected (i.e., leading to an expected flux of
$\approx 5\times 10^{-15}$ erg s$^{-1}$ cm$^{-2}$).  In either case,
our estimate of $\theta_j\gtrsim 8^\circ$ will not change
significantly.

We also detect the host galaxy of \grb, which exhibits red
optical/near-IR colors and radio emission that are well-matched by the
SED of a ULIRG (Arp\,220) at $z\approx 1.3$.  The inferred total star
formation rate is $\approx 300$ M$_\odot$ yr$^{-1}$.  A low-luminosity
radio-loud AGN in an elliptical galaxy cannot be definitively ruled
out, but the ULIRG interpretation, combined with the small projected
offset of $2.2\pm 1.2$ kpc, more naturally explains the inferred
afterglow extinction.  The host galaxy is part of an
interacting/merging system, which is not unexpected for ULIRGs.
Observations with the Atacama Large Millimeter/submillimeter Array
(ALMA) will be able to robustly distinguish the two scenarios.  At 240
GHz (ALMA band 6), the expected flux densities are about 0.3 mJy and 1
$\mu$Jy, for the ULIRG\footnotemark\footnotetext{This flux density is
relevant for the scaled SED of Arp\,220.  Variations in the dust
temperature would lead to a range of potential flux densities (e.g.,
\citealt{yc02,mhc+08}).}  and AGN cases, respectively.  The former can
be detected with high significance in a short observation.  Similarly,
the 60 ks \chandra\ observation at $\delta t\approx 46$ d can achieve
a limiting flux of $F_X\approx 1.5\times 10^{-15}$ erg s$^{-1}$
cm$^{-2}$, which is still about 3 times higher than the expected X-ray
emission due to star formation.  Therefore, if flattening of the X-ray
light curve is detected in this observation, it may point to an AGN
contribution.

The host galaxy of the short GRB\,100206A at $z=0.407$ is also a
luminous infrared galaxy, $L_{\rm IR}\approx 4\times 10^{11}$
L$_\odot$, with ${\rm SFR}\approx 30-40$ M$_\odot$ yr$^{-1}$ inferred
from SED fitting of data at $\sim 0.3-10$ $\mu$m \citep{pmm+11}.  Here
we find ${\rm SFR}\lesssim 60$ M$_\odot$ yr$^{-1}$ from radio
observations.  Of the $\approx 25$ short GRBs with robust host galaxy
associations (Fong et al.~in preparation), the hosts of GRBs 120804A
and 100206A are the only galaxies with clear LIRG/ULIRG properties.
At $z\sim 1$, the population of ULIRGs and bright LIRGs accounts for
$\sim 25\%$ of the total star formation rate density
\citep{lpd+05,cly+07}, but their contribution to the stellar mass
density is small, $\sim 1-{\rm few}\%$ (e.g., \citealt{cdl+06}).
Thus, the fraction of $\sim 5-10\%$ in the short GRB host population
is intermediate between the star formation and mass weighted
fractions.  This is expected for a progenitor population that tracks
both star formation and stellar mass, consistent with previous
findings for short GRBs \citep{lb10}.

\acknowledgments We thank Karina Caputi, Ranga Chary, Francesca
Civano, Martin Elvis, and Jane Rigby for useful discussions about
ULIRGs and AGN.  We also thank Brian Metzger for information on the
magnetar scenario.  The Berger GRB group at Harvard is supported by
the National Science Foundation under Grant AST-1107973, and by
NASA/Swift AO7 grant NNX12AD69G.  E.B.~acknowledges partial support of
this research while in residence at the Kavli Institute for
Theoretical Physics under National Science Foundation Grant
PHY11-25915.  The Dark Cosmology Centre is funded by the Danish
National Research Foundation.  V.~M.~acknowledges funding through
contract ASI-INAF I/004/11/0.  Observations were obtained with the
JVLA (program 12A-394) operated by the National Radio Astronomy
Observatory, a facility of the National Science Foundation operated
under cooperative agreement by Associated Universities, Inc.  The
paper includes data gathered with the 6.5 meter Magellan Telescopes
located at Las Campanas Observatory, Chile; with ESO Telescopes at the
La Silla Paranal Observatory under programme ID 089.D-0450; with the
Gemini Observatory, which is operated by the Association of
Universities for Research in Astronomy, Inc., under a cooperative
agreement with the NSF on behalf of the Gemini partnership: the
National Science Foundation (United States), the Science and
Technology Facilities Council (United Kingdom), the National Research
Council (Canada), CONICYT (Chile), the Australian Research Council
(Australia), Ministério da Ciência, Tecnologia e Inovação (Brazil) and
Ministerio de Ciencia, Tecnología e Innovación Productiva (Argentina).
This work is based in part on observations obtained with XMM-Newton,
an ESA science mission with instruments and contributions directly
funded by ESA Member States and the USA (NASA).

{\it Facilities:} \facility{Swift (XRT)}, \facility{CXO (ACIS-S)},
\facility{XMM-Newton} \facility{Gemini:North (GMOS)},
\facility{Magellan (FourStar)}, \facility{VLT (HAWK-I),
\facility{JVLA}}


\clearpage
\begin{deluxetable}{lccc}
\tablecolumns{4}
\tabcolsep0.05in\scriptsize
\tablewidth{0pc}
\tablecaption{JVLA Observations
\label{tab:evla}}
\tablehead{
\colhead {Date}       &
\colhead {$\delta t$} &
\colhead {Frequency}  &
\colhead {$F_\nu$}    \\
\colhead {(UT)}       &
\colhead {(d)}        &
\colhead {(GHz)}      &
\colhead {($\mu$Jy)}  
}
\startdata
2012 Aug  4.97 & 0.93  & 4.9 & $44\pm 8$  \\
               &       & 6.7 & $25\pm 7$  \\
2012 Aug  5.99 & 1.95  & 4.9 & $40\pm 8$  \\
               &       & 6.7 & $26\pm 7$  \\
2012 Aug  8.02 & 3.98  & 4.9 & $45\pm 9$  \\    
               &       & 6.7 & $25\pm 8$  \\
2012 Aug 14.94 & 10.90 & 4.9 & $45\pm 10$ \\ 
               &       & 6.7 & $24\pm 9$  \\
2012 Sep 11.01 & 37.97 & 4.9 & $\lesssim 40\,^a$ \\
               &       & 6.7 & $\lesssim 36\,^a$ 
\enddata
\tablecomments{$^a$ Limits are $3\sigma$.}
\end{deluxetable}

\clearpage
\begin{deluxetable}{lccccc}
\tablecolumns{6}
\tabcolsep0.08in\scriptsize
\tablewidth{0pc}
\tablecaption{Optical and Near-IR Observations of the Host and Nearby
Galaxy
\label{tab:host}}
\tablehead{
\colhead {Filter}                 &
\colhead {$\lambda_{\rm obs}$}    &
\multicolumn{2}{c}{Host Galaxy}   &
\multicolumn{2}{c}{Nearby Galaxy} \\
\colhead {}                    &
\colhead {}                    &
\colhead {AB Magnitude$\,^a$}  &
\colhead {$F_\nu\,^a$}         &
\colhead {AB Magnitude$\,^a$}  &
\colhead {$F_\nu\,^a$}         \\
\colhead {}                    &
\colhead {($\mu$m)}            &
\colhead {}                    &
\colhead {($\mu$Jy)}           &
\colhead {}                    &
\colhead {($\mu$Jy)}  
}
\startdata
$r$   & 0.630 & $25.9\pm 0.2\,^b$ & $0.16\pm 0.04\,^b$ & $26.0\pm 0.2$  & $0.15\pm 0.03$  \\ 
$i$   & 0.779 & $24.8\pm 0.15$    & $0.44\pm 0.07$     & $24.4\pm 0.1$  & $0.63\pm 0.06$  \\
$Y$   & 1.021 & $23.7\pm 0.3$     & $1.20\pm 0.40$     & $23.0\pm 0.2$  & $2.30\pm 0.46$  \\
$J$   & 1.235 & $23.0\pm 0.2$     & $2.19\pm 0.44$     & $22.4\pm 0.15$ & $3.98\pm 0.60$  \\
$K_s$ & 2.146 & $22.0\pm 0.1$     & $5.75\pm 0.58$     & $21.8\pm 0.1$  & $6.85\pm 0.69$ 
\enddata
\tablecomments{$^a$ These magnitudes and flux densities have been
corrected for Galactic extinction of $E(B-V)\approx 0.204$ mag
\citep{sf11}. $^b$ This value has been corrected for an expected
afterglow contribution of $\approx 0.03$ $\mu$Jy.}
\end{deluxetable}

\clearpage 

\clearpage

\clearpage

\end{document}